# Only Six Passive Circuit Elements Are Existent

Frank Zhigang Wang

Division of Computing, Engineering and Mathematical Sciences, University of Kent, Canterbury CT2 7NF, UK

Corresponding author: Frank Zhigang Wang (E-mail: frankwang@ieee.org).

This research was partially conducted in an EC grant “Re-discover a periodic table of elementary circuit elements”, PIIFGA2012332059, Marie Curie Fellow: Leon Chua (UC Berkeley), Scientist-in-charge: Frank Wang (University of Kent).

***Abstract*—We found that a second-order ideal memristor degenerates into a negative nonlinear resistor. This phenomenon is quite similar to what are observed in chemistry: a chemical element with a higher atomic number is unstable and may decay radioactively into another chemical element with a lower atomic number. After extending the above local activity (verified both graphically and analytically) to other higher-order circuit elements, we concluded that all higher-order passive memory circuit elements do not exist in nature and that the periodic table of the two-terminal passive circuit elements can be dramatically reduced to a six-pointed star comprising only six passive elements. Such a bounded table may mark the end of the hunt for missing higher-order passive circuit elements predicted 40 years ago.**

## I. Introduction

In 1982, Chua and Szeto introduced higher-order and mixed-order nonlinear circuit elements in a periodic table of all two-terminal nonlinear circuit elements ($\alpha$, $\beta$) that is periodic modulo ±4 [1], as shown in Fig.1. They presented that a two-terminal circuit element characterized by a constitutive relation in the $v^{(\alpha)}$-versus-$i^{(\beta)}$ plane is called an ($\alpha$, $\beta$) element, where $v^{(\alpha)}$ and $i^{(\beta)}$ are complementary variables derived from the voltage $v(t)$ and current $i(t)$, respectively [1]. $\alpha$ and $\beta$ can be any positive integer (representing $\alpha th$-order or $\beta th$-order differential wrt time), any negative integer (representing $\alpha th$-order or $\beta th$-order integral wrt time), or zero [1]. They demonstrated that most of the circuit elements in the table are active and not lossless [1]. They predicted that the only nonlinear candidates for passivity are the elements that lie on the three diagonals (in red), as shown in Fig.1 [1].

In 1984, Chua and Szeto further demonstrated that any higher nonlinear n-port element with a defined constitutive relation can be realized using a mixture of linear elements, nonlinear elements and power sources [2].

In 2003, Chua picked resistor ($\alpha=0$, $\beta=0$), inductor ($-1$, $0$), capacitor ($0$, $-1$) and memristor (short for memory resistor) ($-1$, $-1$) [3][4][5] to form a four-element torus and used it as a seed to generate all other ($\alpha$, $\beta$) elements [3], including higher-order memory circuit elements [memristor ($\alpha\leq-2$, $\beta\leq-2$), mem-inductor ($\alpha\leq-3$, $\beta\leq-2$), and mem-capacitor ($\alpha\leq-2$, $\beta\leq-3$).

In 2009, Biolek, Biolek and Biolkova defined second-order devices by introducing a pair of new integral variables, namely, $\sigma=\int q dt$, $\rho=\int \varphi dt$, where $\varphi$ is the magnetic flux and $q$ is the electric charge [6].

In 2009, Pershin, Ventra and Chua studied charge-controlled memcapacitors and flux-controlled meminductors [7].

In 2013, Riaza demonstrated that the corresponding second order mem-circuits based on those second-order elements yield rich dynamic behavior (e.g., bifurcation phenomena) [8].

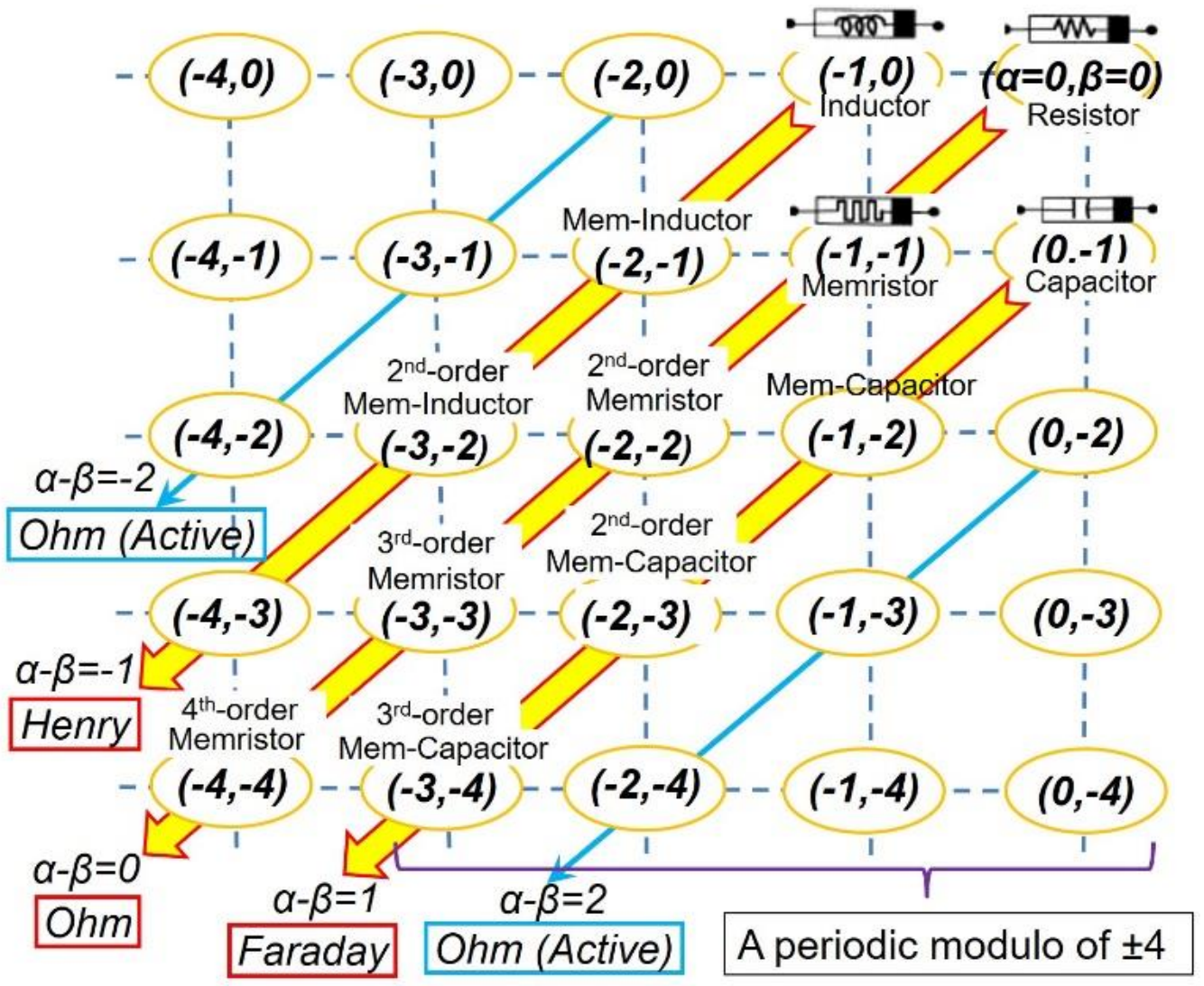

*Fig.1 Chua’s periodic table of all two-terminal nonlinear circuit elements (α, β) that is periodic modulo ±4. This table is unbounded since Chua thought he could use his four-element torus (resistor, inductor, capacitor and memristor) as a seed to generate all other elements endlessly [3]. Chua and Szeto predicted that most of the higher-order circuit elements are active and the only passive nonlinear candidates are on the three diagonals (in red) [1].*

In 2012, Chua, et. al, demonstrated that the time-varying potassium conductance and the time-varying sodium conductance in the Hodgkin-Huxley circuit are first-order potassium memristor and second-order sodium memristor, respectively [9]. Both memristors are locally active with an internal battery [9][10].

In 2019, Biolek, Biolek and Biolkova proved that Hamilton’s variational principle still holds true for circuits consisting of ($\alpha$, $\beta$) elements from Chua’s periodical table [11].

Our study is the first work inspired by the fact that, in Mendeleev’s periodic table of chemical elements, chemical elements beyond uranium (with atomic number 92) were not found in nature because they were too unstable and all chemical elements with a higher atomic number than 118 remain purely hypothetical [12]. We study whether Chua’s predicted higher-order passive circuit element candidates on the three diagonals in his periodic table [1] exist or not.

## II. A First-Order Ideal Memristor Is Locally Passive

For the information integrity, this section is intended to list all the relevant facts and prepare some necessary prerequisites for the following sections although it is common knowledge that a first-order ideal memristor is locally passive.

First-order circuit elements are those defined by their constitutive attributes $\varphi$ and $q$ [1][2][3][4][5]. The memristor, with memristance *M*, provides a constitutive relation between the charge $q$ and the flux $\varphi$ as given under $d\varphi/dq = M(q)$ memristance. The conformal differential transformation [3] from the original $\varphi$-$q$ plane to the *v*-*i* plane is shown in Fig.2.

As shown in Fig.2, an arbitrary $\varphi = \hat{\varphi}(q)$ curve in the $\varphi$-$q$ plane represents a generic memristor. This curve should be origin-crossing as $\hat{\varphi}(q) = M(q) \cdot q$. Note that the memristor shown in Fig.2 is charge-controlled but the principles found here should be applicable to another type of flux-controlled memristor. In Fig.2, $\hat{\varphi}(q)$ and $\hat{v}(i)$ are continuous and piecewise-differentiable functions with bounded slopes [5]. Note that the voltage *v* is a function not only of *i* but also of *q*; therefore, $\hat{v}(i)$ should be a double-valued function of the current *i* because every pair of points $A(t_0)$ and $B(\pi\text{-}t_0)$ on the $\hat{\varphi}(q)$ curve (that are subject to the same amplitude of an excitement current) are separated vertically by following their own chord slopes in the *v*-*i* plane (to be detailed later).

It is convenient to assume a cosine charge function to cover the full operating range {0, 2}, as used by Chua [4]. Note that full-range scanning is necessary to expose a distinctive "fingerprint"; otherwise, the obtained fingerprint is incomplete. This function is defined by

$$\begin{cases} q(t) = 1 - \cos t, & t \geq 0 \\ \quad\quad\; = 0, & t < 0 \end{cases}$$

where the initial charge $q(0) = 0$.

Its corresponding current, *i(t)*, as a testing signal across the memristor, is:

$$\begin{cases} i(t) = \sin t, & t \geq 0 \\ \quad\;\; = 0, & t < 0 \end{cases}$$

There must be a correspondence between the $\varphi$–$q$ plane and the *v*-*i* plane because the coordinates of the latter are the differential (with respect to time) of the coordinates of the former [3]. It is easy to have:

$$\alpha = \tan^{-1}\frac{d\varphi(t)}{dq(t)} = \tan^{-1}\frac{d\varphi(t)/dt}{dq(t)/dt} = \tan^{-1}\frac{v(t)}{i(t)} = \alpha'. \qquad (1)$$

That is, the slope ($\alpha$) of the line tangent to the $\varphi = \hat{\varphi}(q)$ curve at an operating point $A(t{=}t_0)$ in the $\varphi$–$q$ plane is equal to the slope ($\alpha'$) of a straight line connecting the projected point $A'(t{=}t_0)$ to the origin in the $v = \hat{v}(i)$ plane (on the same scale as the $\varphi$–$q$ plane) [5].

This conformal differential transformation can be simplified as follows: Linearizing $\varphi = \hat{\varphi}(q)$ at operating point $A(\varphi_0, q_0)$ at time $t_0$ (Fig.2) via series expansion, $\varphi = \varphi_0 + \left.\frac{d\hat{\varphi}}{dq}\right|_{q_0}(q - q_0)$ is obtained which is the equation for the tangent at $A(\varphi_0, q_0)$. Differentiating the above equation for the tangent wrt time $v(t) = \left.\frac{d\hat{\varphi}}{dq}\right|_{t_0} i(t)$ is obtained which is the equation of the line joining the origin to the projected operating point *A'* at $t_0$ in the *v*-*i* plane. So the slopes match.

A simple graphic method to draw the voltage-current loci $v = \hat{v}(i)$ corresponding to the above given $\varphi = \hat{\varphi}(q)$ curve is illustrated in Fig.2: (1) Obtaining $\alpha$ at an operating point $A(t{=}t_0)$ in the $\varphi = \hat{\varphi}(q)$ curve, one draws a straight line through the origin in the *v*-*i* plane whose slope is $\alpha' = \alpha$; (2) Projecting point $A(t{=}t_0)$ from the $\varphi$–$q$ plane onto the *v*-*i* plane by following Projection Lines ①, ② and ③, one eventually ends up with the same time point $A'(t{=}t_0)$ in the *v*-*i* plane by meeting Projection Line ③ with the drawn line in the first step.

As shown in Fig.2, an arbitrary $\varphi = \hat{\varphi}(q)$ curve results in a generic memristor meeting the distinctive "fingerprint" [4]: 1. Zero-crossing or pinched ($\because v = M(q)i, \therefore v = 0$ whenever $i{=}0$); 2. Double-valued Lissajous figure.

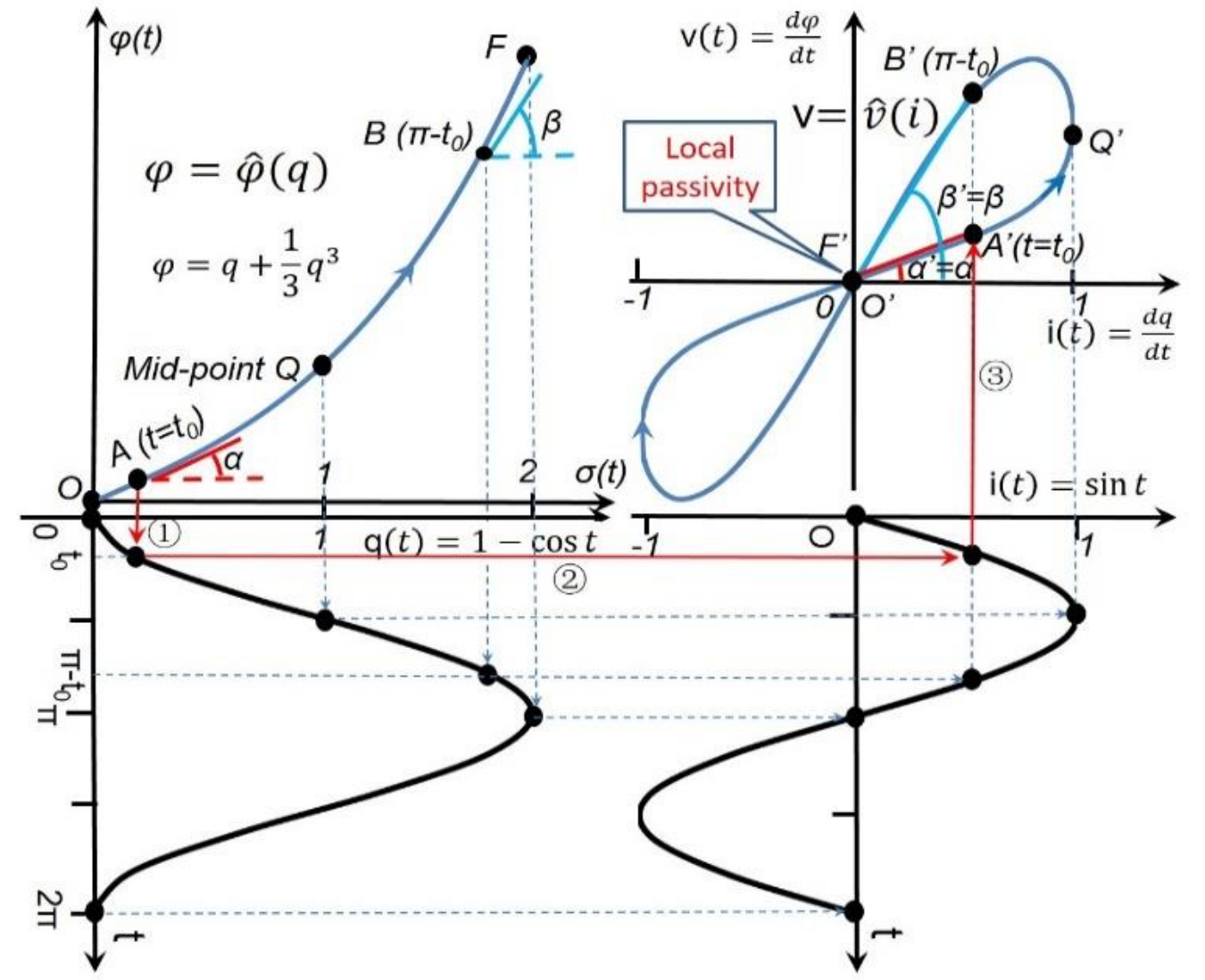

*Fig.2 First-order ideal memristor with a single-valued, differentiable, monotonically increasing $\hat{\varphi}(q)$ curve and a odd-symmetric pinched v-i hysteresis loop against the origin. Based on a "conformal differential transformation" that preserves angles between $\varphi = \hat{\varphi}(q)$ and $v = \hat{v}(i)$, projection lines 1, 2 and 3 denote a simple graphic method to project a given $\hat{\varphi}(q)$ curve in the $\varphi$–q plane onto the v-i plane. A' denotes A's projected point after one transformation, and so on. A small region (e.g., the short arc above Q') with a negative slope does not imply that there is an internal power source inside this first-order memristor. Local passivity [13] is validated by the origin-crossing of the v-i loci rather than the slope of any operating point on the v-i loop (see main text for details).*

A cell is said to be locally passive at a cell equilibrium point if, and only if, it is not locally active at that point [13].

If the $\varphi = \hat{\varphi}(q)$ curve is split into two branches (the outgoing path does not overlap the returning path), the pinched figure $v = \hat{v}(i)$ becomes asymmetric with respect to the origin, as shown in Fig.3. This represents a wide range of practical nonideal memristors with an asymmetric bipolar or unipolar pinched Lissajous figure [14].

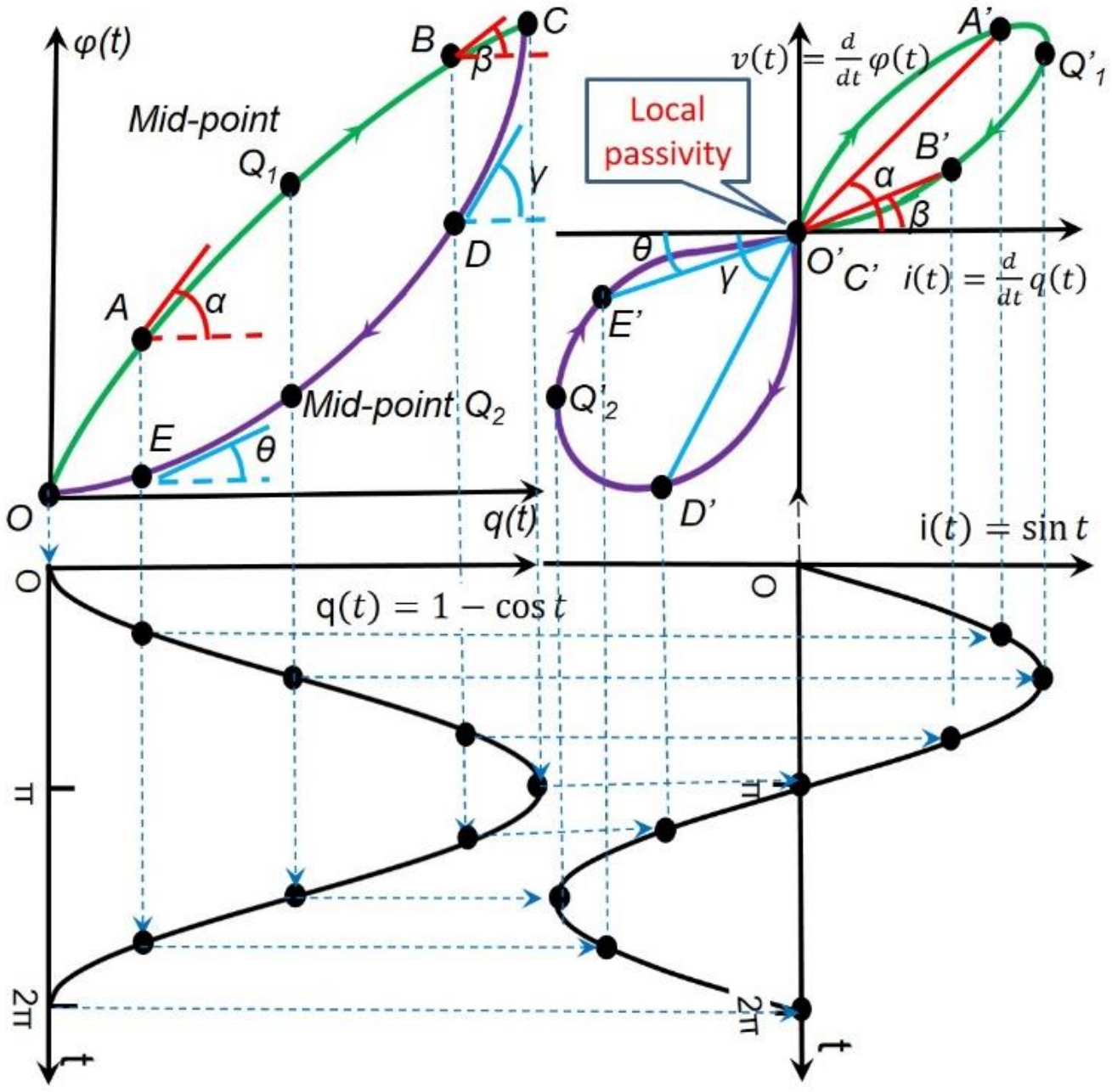

*Fig.3 A nonideal memristor with two φ-q characteristic branches (in different colors) and an asymmetric pinched v-i hysteresis loop. Note that each branch of the φ-q curve generates only one loop of the Lissajous figure $\hat{v}(i)$. Local passivity is validated by the origin-crossing of the v-i loci.*

An ideal first-order memristor should be characterized by a time-invariant *φ-q* curve complying with the following four criteria:

1. Single-valued*;
2. Nonlinear;
3. Continuously differentiable;
4. Strictly monotonically increasing.

Criteria 2-4 for ideality were extracted from references [4][14]. The only exception is that we added the "single-valued" criterion. Strictly speaking, if it is strictly monotonic increasing, it is single-valued. *Listing the "single-valued" criterion separately is to clearly exclude the special case (Fig.3), in which a nonideal memristor has a doubled-valued, strictly-monotonic-increasing φ-*q* curve.

The following passivity theorem for an ideal first-order memristor is therefore reasoned:

| Theorem I: Local Passivity Theorem |
|---|
| A first-order ideal memristor is locally passive. |

Proof:
Criteria 4 (strictly monotonically increasing) means that, if *A*<*B*, f(*A*)<f(*B*) by monotonicity, thus the slope of $\hat{\varphi}(q)$ is nonnegative; hence, this ideal memrisor is locally passive at each point on the *φ–q* curve.

This theorem is important for small-signal circuit analysis since a locally active memristor may give rise to oscillations, and even chaos [4]. On the other hand, such passivity leads to energy efficiency in green computing because one does not have to keep supplying power to make a passive circuit element function (In contrast, an active-transistor-based memory cell will lose its stored information after the power supply is switched off.).

A careful examination of the $v = \hat{v}(i)$ loci in Fig.2 reveals that the pinched hysteresis loop contains a small region (e.g., the short arc *above Q'*) with a negative slope, which does not imply that there is an internal power source inside this first-order memristor. Such a negative slope appears because of the phase lag between the peak of the voltage waveform and the peak of the current waveform [1]. As shown in Fig.2, the aforementioned local passivity is validated by the origin-crossing of the *v-i* loci rather than the slope of any operating point on the *v-i* loop.

Such a local passivity should be applicable to both first-order mem-inductors and mem-capacitors. The proof is omitted here due to similarity and simplicity. The reader should repeat a deduction, similar to that used in Fig.2.

## III. A Second-Order or Higher-order Ideal Memristor Degenerates into a Negative Nonlinear Resistor

In this section, we use a second-order memristor ($\alpha$=-2, $\beta$=-2) as an example to investigate whether it has to be locally active, as demonstrated in Chua's emulation of the Hodgkin-Huxley circuit [9][10].

Beyond the first-order setting, second-order circuit elements require double-time integrals of voltage and current, namely, $\sigma = \int qdt = \iint idt$ and $\rho = \int \varphi dt = \iint vdt$. With the use of these additional variables, we may accommodate a second-order memristor and its nonlinear counterparts to be described in the following sections.

In the same way to define a first-order ideal memristor, a second-order ideal memristor should be characterized by a time-invariant *ρ-σ* curve (Fig.4) complying with the same set of four criteria as mentioned above.

With the aid of the aforementioned conformal differential transformation, such a $\rho = \hat{\rho}(\sigma)$ curve in the constitutive *ρ-σ* plane is transformed into a zero-crossing, double-valued Lissajous figure $\varphi = \hat{\varphi}(q)$ in the *φ-q* plane. As shown in Fig.4, following the *O-Q-A-B-F* path during the first half cycle ($0 \le t \le \pi$) in the *ρ-σ* plane, the chord (a straight line connecting a point to the origin) sweeps in a counterclockwise direction in the *O'-Q'-A'-B'-F'* order in the first quadrant of the *φ-q* plane, and then repeats the sweep in an asymmetric manner in the third quadrant during the second half cycle ($\pi \le t \le 2\pi$), resulting in an anti-symmetric pinched hysteresis loop $\varphi = \hat{\varphi}(q)$. Because the *ρ-σ* curve is single-valued, the outgoing path ($0 \le t \le \pi$) overlaps the returning path ($\pi \le t \le 2\pi$) in $\rho = \hat{\rho}(\sigma)$ and the corresponding pinched Lissajous figure $\varphi = \hat{\varphi}(q)$ is anti-symmetric with respect to the origin of the *φ-q* plane.

Consequently, another conformal differential transformation is performed, resulting in the slope ($\theta$) of the line tangent to the $\varphi = \hat{\varphi}(q)$ curve at an operating point *A'*($t$=$t_0$) in the *φ–q* plane being equal to the same slope ($\theta$) of a straight line connecting the projected point *A''*($t$=$t_0$) to the origin in the *v-i* plane. Because the *φ–q* curve is anti-symmetric with respect to its midpoint of the operating range (in this case, it is the origin), the corresponding *v-i* locus is single-valued, as shown in Fig.4.

The mathematical proof for being a single-valued function is as follows.

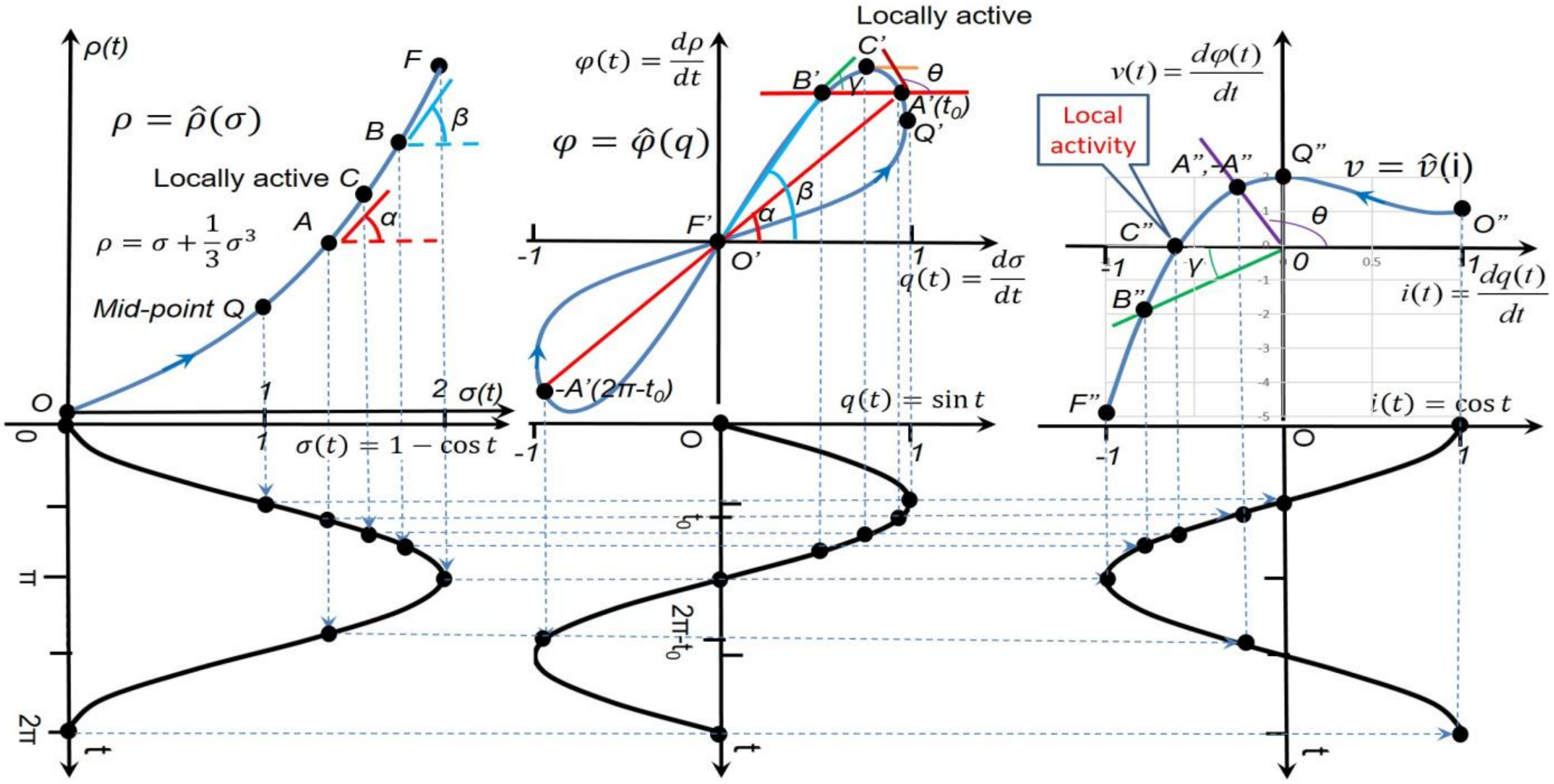

*Fig.4 A second-order ideal memristor requires double-time integrals of voltage and current, namely, σ=∫qdt=∬idt and ρ=∫φdt=∬vdt, as its constitutive attributes. The constitutive ρ-σ curve should be time-invariant, single-valued, nonlinear, differentiable, and monotonically increasing to ensure its ideality. Two consecutive conformal differential transformations are required to generate v-i locus that exhibits local activity and manifests itself as a negative nonlinear resistor. A'' denotes A's projected point after two transformations. A polynomial* $\rho = \sigma + \frac{1}{3}\sigma^3$ *is used here as an example to draw this graph without losing generality since the proof in the main text is carried out in a generic case.*

By the definition of anti-symmetry [$f(-y, -x) = -f(x, y)$] and $q(t_0) = \sin t_0 = -\sin(2\pi - t_0) = -q(2\pi - t_0)$, $i(t_0) = \cos t_0 = \cos(2\pi - t_0) = i(2\pi - t_0)$, we have

$$\varphi(t_0) = -\varphi(2\pi - t_0) \tag{2}$$

for a pair of points $A'(t_0)$ and $-A'(2\pi - t_0)$ that are anti-symmetric against the midpoint of the $\varphi$–$q$ loop (overlapping the origin of the $\varphi$–$q$ plane in Fig.4). We then obtain

$$\left.\frac{d\hat{\varphi}(q)}{dq}\right|_{t=t_0} = \left.\frac{d\hat{\varphi}(q)}{dq}\right|_{t=2\pi-t_0}, \tag{3}$$

and

$$v(t_0) = \left.\frac{d\varphi(t)}{dt}\right|_{t=t_0} = \left.\frac{d\hat{\varphi}(q)}{dq(t)}\frac{dq(t)}{dt}\right|_{t=t_0} = \left.\frac{d\hat{\varphi}(q)}{dq(t)}\right|_{t=t_0} \cdot i(t_0) = \left.\frac{d\hat{\varphi}(q)}{dq(t)}\right|_{t=2\pi-t_0} \cdot i(2\pi - t_0) = \left.\frac{d\varphi(t)}{dq(t)}\frac{dq(t)}{dt}\right|_{t=2\pi-t_0} = \left.\frac{d\varphi(t)}{dt}\right|_{t=2\pi-t_0} = v(2\pi - t_0). \tag{4}$$

The above two anti-symmetric points $A'(t_0)$ and $-A'(2\pi - t_0)$ in the $\varphi$–$q$ plane are projected onto the *v*-*i* plane as one single point *A*''(-*A*''); therefore, the $\hat{v}(i)$ loci collapse into another single-valued function without passing through the origin. The above deduction can be summarized as:

> Theorem II: Single Value Theorem
> A single-valued function becomes another single-valued function after two consecutive conformal differential transformations.

The circuit element with such a single-valued *v*-*i* curve (as shown in Fig.4) that does not go through the origin manifests itself as a negative nonlinear resistor [3], in which there must be an internal power source (either a current source or a voltage source). On the other hand, this single value theorem ensures the zeroth-order of a negative nonlinear resistor, into which a second-order memristor degenerates.

An examination of the $\varphi = \hat{\varphi}(q)$ loci in Fig.4 reveals that this pinched hysteresis loop contains a small region (e.g., the short arc *C'A'Q'*) with a negative slope, which results in the projected points *A''* being located in the second quadrant of the *v*-*i* plane. This observation also leads to the same conclusion that a second-order ideal memristor is locally active.

The mathematical proof of the passivity is as follows.

As shown in Fig.5, the mean value theorem [15] in mathematics states that if *f* is continuous on the closed interval [a, b] and differentiable on the open interval (a, b), then there must be a point c in (a, b) such that the tangent at c is parallel to the secant line joining the endpoints (a, f (a)) and (b, f (b)), that is,

$$f'(c) = \frac{f(b) - f(a)}{b - a}. \tag{5}$$

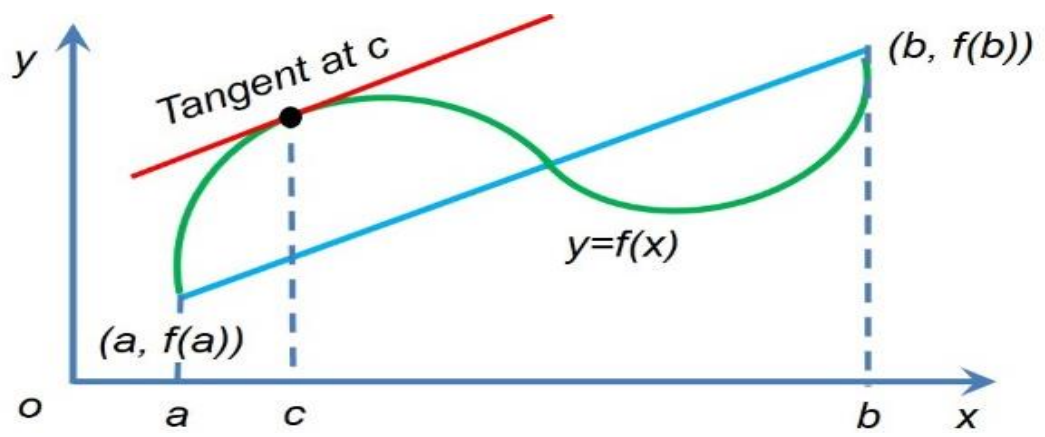

*Fig.5 Cauchy's mean value theorem [15]. For any function that is continuous and differentiable, a point c on (a,b) must exist such that the secant joining the endpoints of [a,b] is parallel to the tangent at c.*

Now let us go back to Fig.4. As proven in Section II, the $\varphi = \hat{\varphi}(q)$ curve is a pinched Lissajous figure. Therefore, we can draw a horizontal secant to have two intersecting points *A'* and *B'* with the $\hat{\varphi}(q)$ loop, as shown in Fig.4. According to the mean value theorem, a point *C'* in (*A', B'*) must exist such that the tangent at *C'* is zero. With the aid of the second conformal differential transformation, *C'* is projected onto the *v-i* plane as *C''*, which is located on the *i* axis and apart from the origin ($C'' \neq O''$). This is typically local activity [13].

Point *C*' is so important to expose local activity. The origination of *C*' in Fig.4 is illustrated in Fig.6, which corresponds to a maxima ($\frac{d\varphi}{dt} = 0$) of *φ(t)*. The phase lag between the peak of the *φ(t)* waveform and the peak of the *q(t)* waveform can be clearly seen in this figure, which generates a small region (*Q', C'*) with a negative slope in the *φ-q* plane. The end-point *Q*' is projected from the mid-point *Q* of the constitutive *ρ-σ* curve as shown in Fig.4.

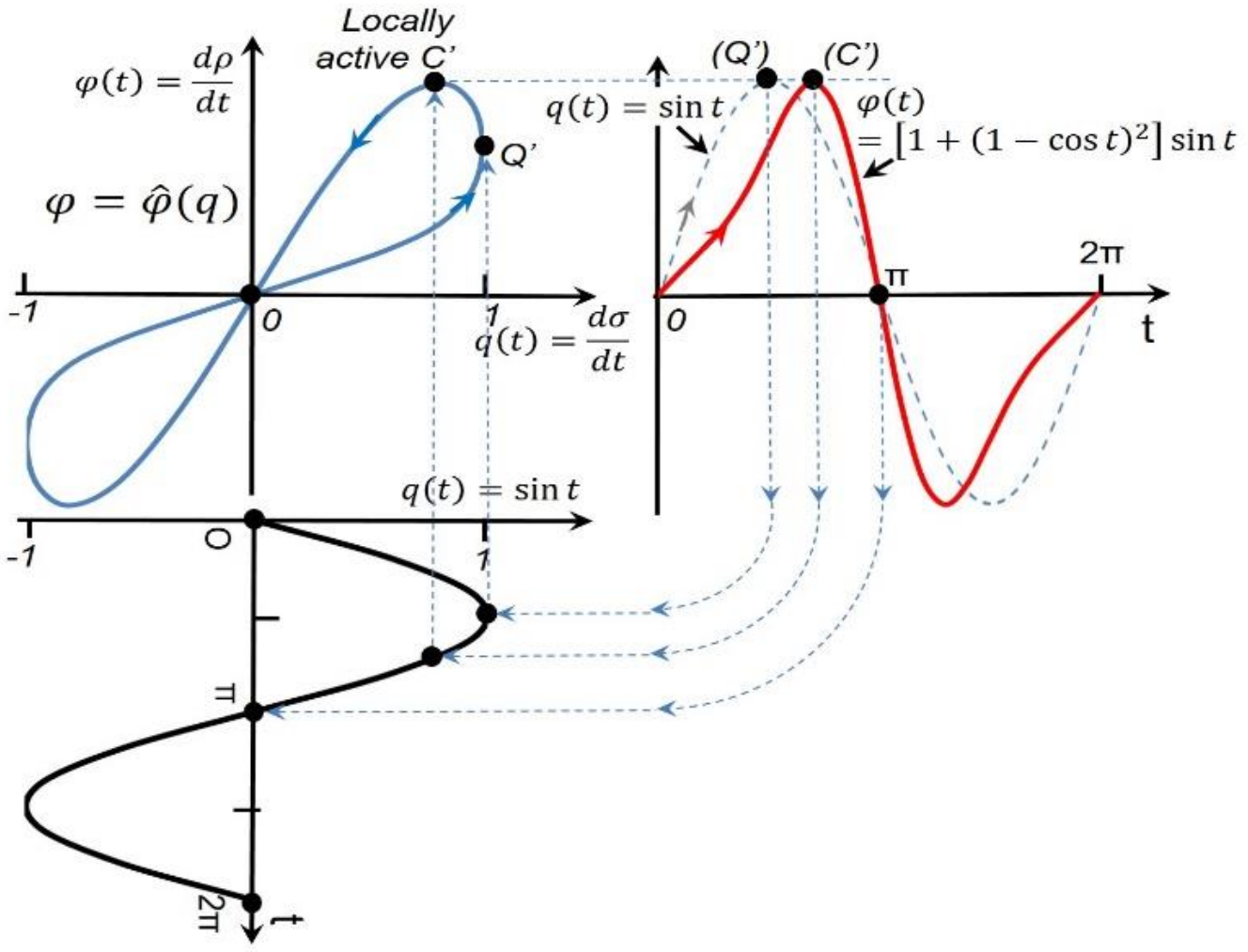

*Fig.6 The local activity point C' in Fig.4 [$\rho = \sigma + \frac{1}{3}\sigma^3$] is originated from the peak of φ(t), which lags behind the peak of q(t), generating a small region (Q', C') with a negative slope. Since $\sigma(t) = 1 - \cos t$, we obtain $\varphi(t) = \frac{d\rho}{dt} = \frac{d(\sigma + \sigma^3/3)}{dt} = \frac{d(\sigma + \sigma^3/3)}{d\sigma}\frac{d\sigma}{dt} = (1 + \sigma^2)\sin t = [1 + (1 - \cos t)^2]\sin t$. To facilitate the phase comparison, the height of the dashed q(t) waveform is scaled up to be aligned with that of φ(t).*

Fig.4 and Fig.6 show an exemplified polynomial $\rho = \sigma + \frac{1}{3}\sigma^3$, which is the same as that used in Chua's tutorial [3]. To a certain extent, this type of polynomial with unbounded slopes means "uncontrolled inflation" between the two variables. Fig.7 uses a logistic function $\rho = \tanh \sigma$, which represents another type of "self-limiting" constitutive relation.

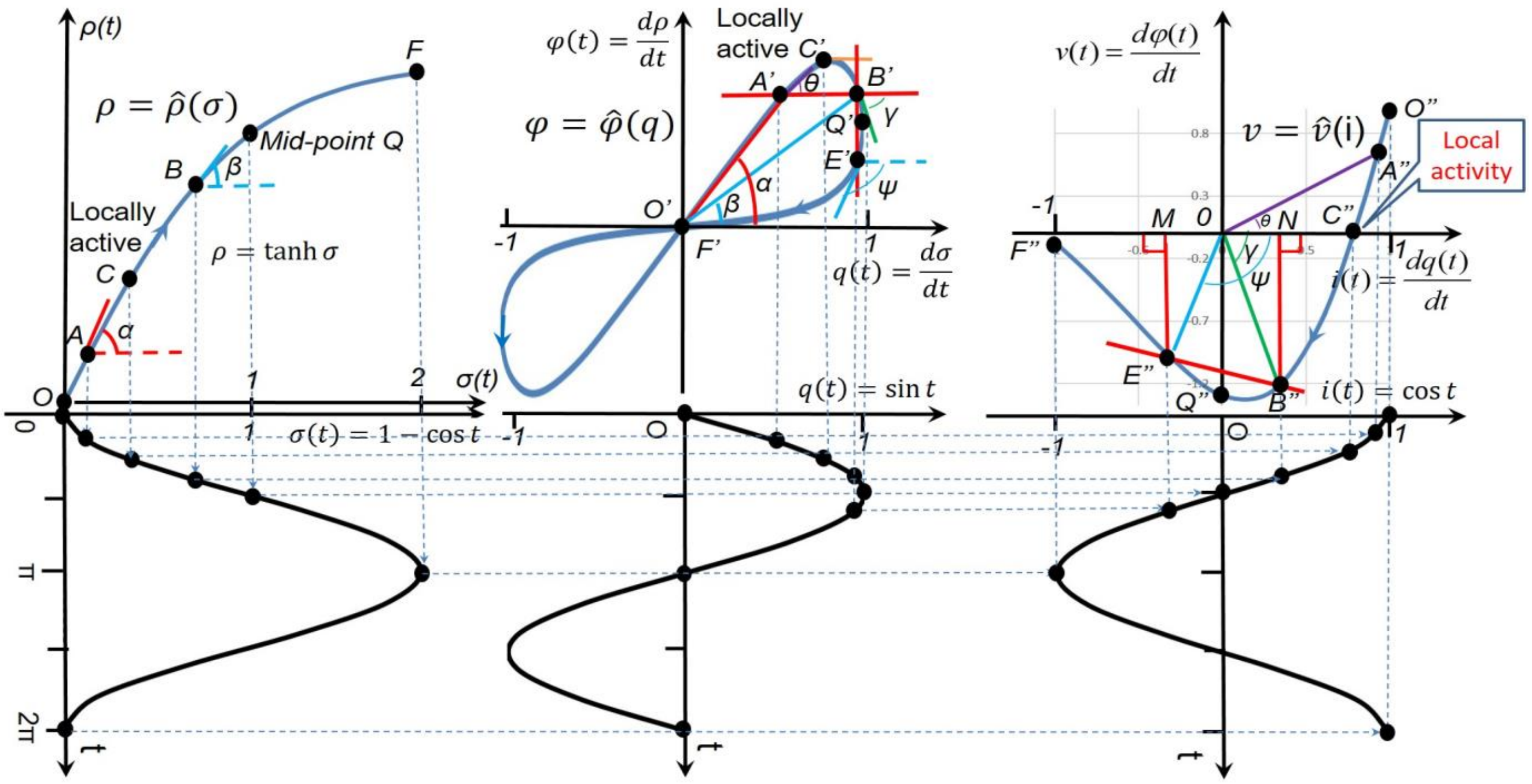

*Fig.7 A logistic function $\rho = \tanh \sigma$ is used to draw the graph for a second-order ideal memristor, representing another type of constitutive relation in terms of self-limiting. Local activity can also be seen after two conformal differential transformations in this case and the second-order memristor still degenerates into a negative nonlinear resistor.*

$\rho = \tanh \sigma$ is a logistic function [16]: $f(x) = \frac{1}{1+e^{-x}} = \frac{1}{2}\tanh\left(\frac{x}{2}\right) + \frac{1}{2}$. For values of $x$ from $-\infty$ to $+\infty$, the $S$-curve shown in Fig.6, with the graph of $f$ approaching 1 as $x$ approaches $+\infty$ and approaching zero as $x$ approaches $-\infty$. The logistic function finds applications in many fields.

In artificial neural networks, the *tanh* function is used as a well-used activation function that is biologically reflected in the neuron (it stays at zero until input current is received, increases the firing frequency quickly at first, but gradually approaches an asymptote at a 100% firing rate).

In biology and ecology, a self-limiting colony of organisms limits its own growth by its actions (releasing waste that is toxic to the colony once it exceeds a certain population) [17]. Fig.7 vividly depicts a self-limiting interaction between $\rho$ and $\sigma$ in a circuit element: $\rho$ responds to $\sigma$ sensitively at the beginning; saturation of $\rho$ occurs while $\sigma$ approaches $+\infty$.

Local activity can also be seen in Fig.7 since the projected point $C''$ is located on the $i$ axis and apart from the origin ($C'' \neq O''$). There must be an internal current source.

A similar examination of the $\varphi = \hat{\varphi}(q)$ loci in Fig.7 reveals that the pinched hysteresis loop contains a small region (e.g., the short arc $C'B'Q'$) with a negative slope, which results in the projected points $B''$ being located in the fourth quadrant of the $v$-$i$ plane. This observation also leads to the same conclusion that a second-order ideal memristor is locally active.

The origination of $C'$ in Fig.7 ($\rho = \tanh \sigma$) is illustrated in Fig.8, which corresponds to a maxima ($d\varphi/dt=0$) of $\varphi(t)$. The phase advance between the peak of the $\varphi(t)$ waveform and the peak of the $q(t)$ waveform can be clearly seen in this figure, which generates a small region ($C'$, $Q'$) with a negative slope in the $\varphi$-$q$ plane.

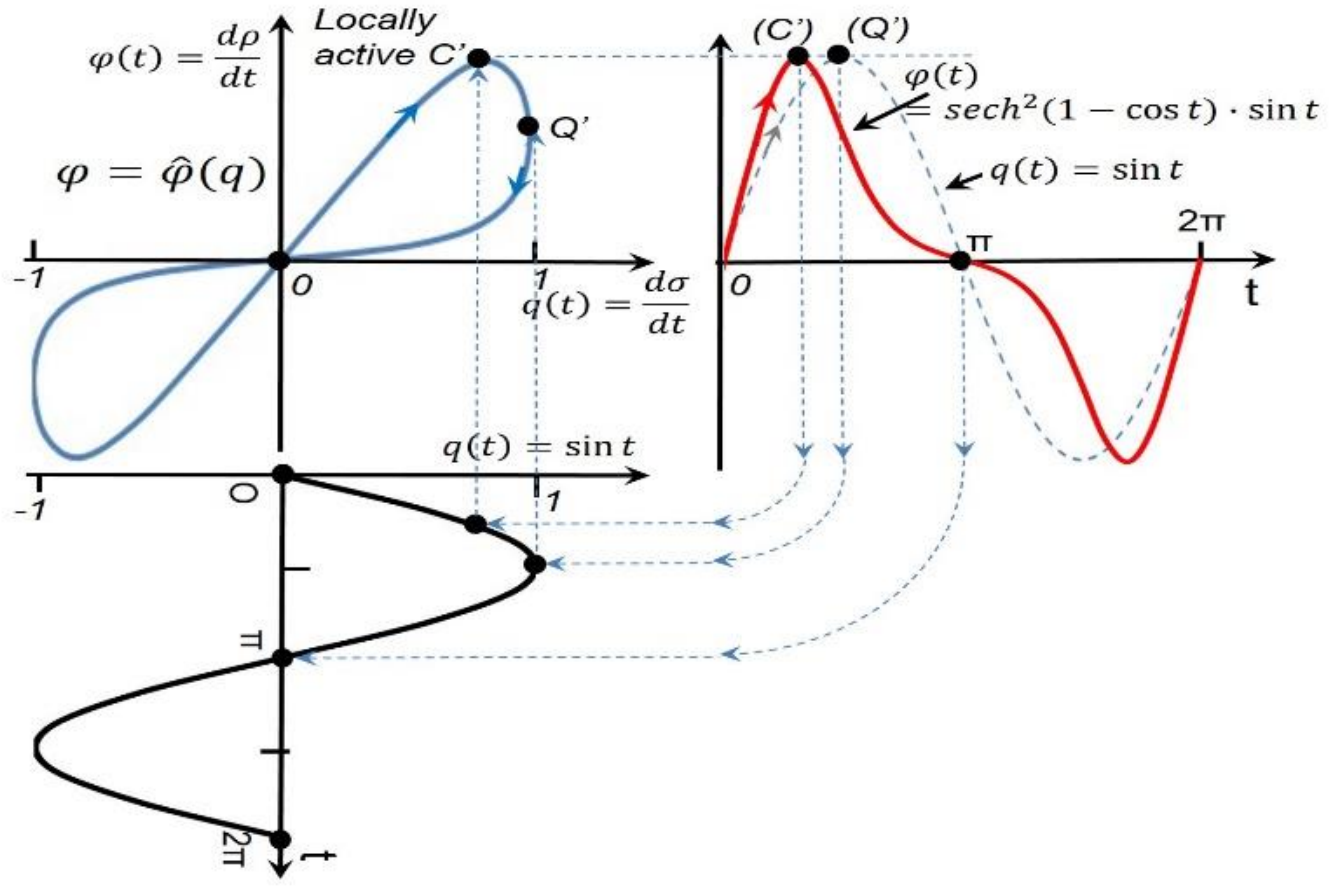

*Fig.8 The local activity point C' in Fig.7 [$\rho = \tanh \sigma$] is originated from the peak of φ(t), which leads the peak of q(t), generating a small region (C', Q') with a negative slope. Since $\sigma(t) = 1 - \cos t$, we obtain $\varphi(t) = \frac{d\rho}{dt} = \frac{d(\tanh \sigma)}{dt} = sech^2(\sigma)\frac{d\sigma}{dt} = sech^2(1-\cos t)\cdot \sin t$. To facilitate the phase comparison, the height of the dashed q(t) waveform is scaled up to be aligned with that of φ(t).*

A careful comparison of Fig.6 and Fig.8 reveals that the concave-convex orientation of the constitutive relation (as displayed in Fig.4 and Fig.7) plays two roles: on the one hand, it determines the winding direction of the $\hat{\varphi}(q)$ loop; on the other hand, the former ($\rho = \sigma + \frac{1}{3}\sigma^3$) causes a phase-lag between $\varphi(t)$ and $q(t)$ [accordingly the local activity point is located in the second-half of the constitutive curve] whereas the latter ($\rho = \tanh \sigma$) causes a phase-advance [accordingly the local activity point is located in the first-half of the constitutive curve].

An even-higher-order memristor ($\alpha \leq -3$, $\beta \leq -3$) can be studied by relaying a sequence of conformal differential transformations from the second-order. As shown in Fig.7, a pair of consecutive points $B'$ and $E'$ in the $\varphi$-$q$ plane ($E'$ leads $B'$) are projected onto the $v$-$i$ plane as $B''$ and $E''$ that are subject to the same amplitude of an excitement current $i$. The height $B''M$ of the laggard point $B''$ is found to be even lower than the height $E''N$ of the leading point $E''$ as $\pi - \psi < \gamma$. The existence of such a pair of points obviously violates the fourth criterion (strictly monotonically increasing) of the memristor ideality stated above. This trajectory makes it impossible for a third-order memristor to be ideal, in which another transformation is needed, in addition to the two transformations shown in Fig.7, to further project a $v$-$i$ curve with a mixture of monotonically increasing and decreasing regions onto a sub-$v$-$i$ plane. Intuitively, even if a single-valued, monotonically increasing constitutive curve cannot guarantee a passive $v$-$i$ curve for a third-order memristor, an even-higher-order one should exhibit a more complicated negative-nonlinear-resistor-like $v$-$i$ behavior.

The following activity theorem for a second-order or higher-order ideal memristor is therefore proven:

| Theorem III: Local Activity Theorem |
|---|
| A second-order or higher-order ideal memristor should be locally active. |

## IV. An Analytical Proof of Local Activity

So far, we have graphically verified the local activity of a second-order or higher-order ideal memristor through the conformal differential transformation [3] and Cauchy's mean value theorem [15]. Next we will use a purely analytical method to verify the aforementioned activity with general constitutive variables ($x$, $y$). An analytical proof begins with a basic axiom or definition (in this case, it is the definition of an ideal memristor) and reach its conclusion through a sequence of deductions and mathematical reasoning. As can be seen later, the so-called analytical method will provide a mathematically different and independent proof from the utilized graphical method in Section III. In scientific research, it is a common practice for a theorem to be established using different or complementary techniques and methods in terms of double-checking (or multi-checking) the conclusion independently.

As shown in the constitutive $x - y$ plane in Fig.9(a), an ideal first-order memristor should be characterized by a time-invariant constitutive curve that is single-valued, nonlinear, continuously differentiable, and strictly monotonically increasing [4][14].

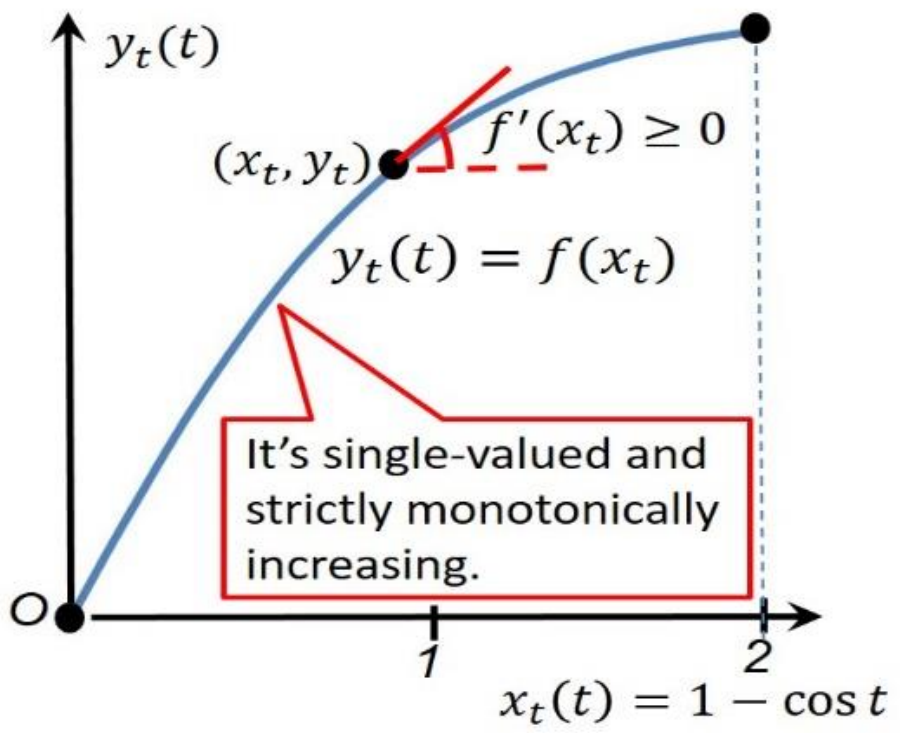

(a) The constitutive $x - y$ plane;

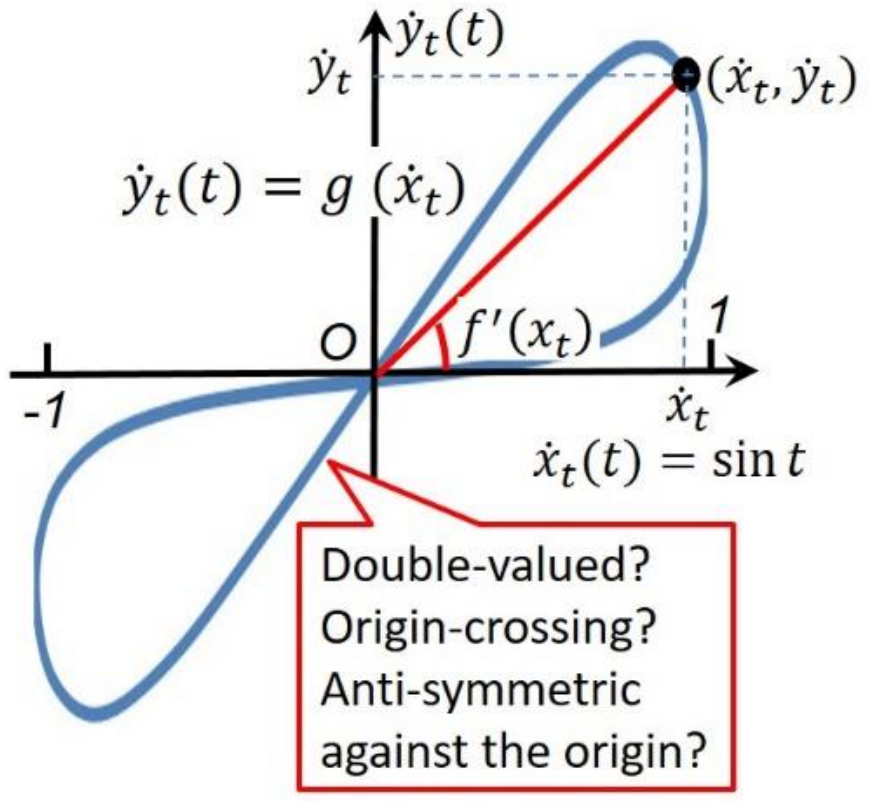

(b) The $\dot{x} - \dot{y}$ plane;

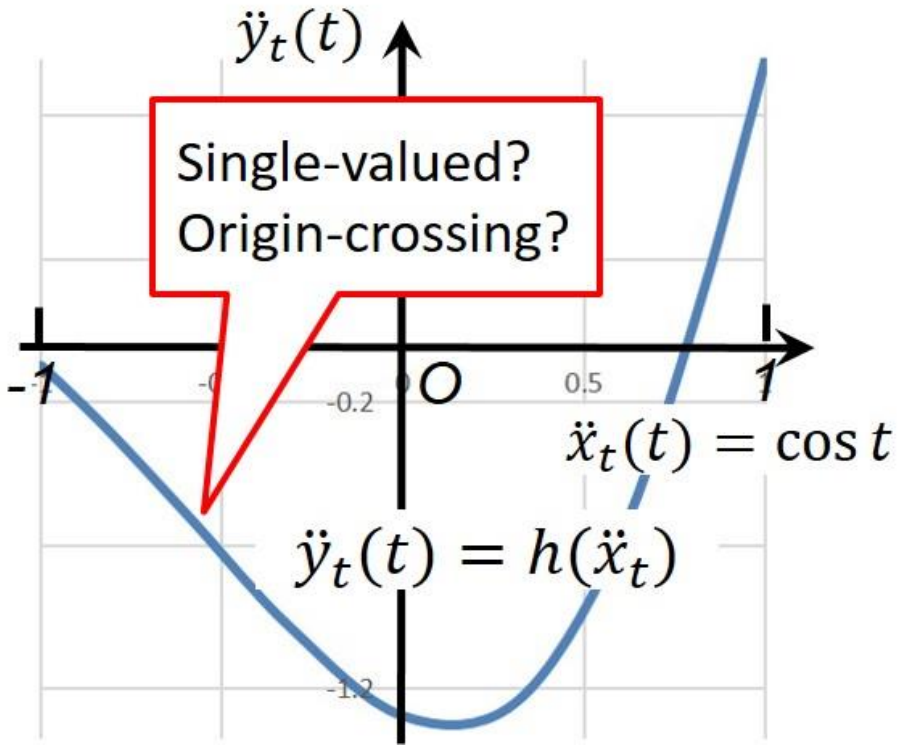

(c) The $\ddot{x} - \ddot{y}$ plane.

*Fig.9 A purely analytical method is used to verify the local activity of a second-order ideal memristor with general constitutive variables (x, y). The use of a subscript t means that $x_t$ is a function of t and so on. The same excitement condition [$x_t(t) = 1 - \cos t$, $\dot{x}_t(t) = \sin t$ and $\ddot{x}_t(t) = \cos t$] as that in Fig.4 and Fig.7 is used without losing the generality of this proof. In mathematics, a function f is called monotonically increasing if for all $x_1$ and $x_2$ such that $x_1 \leq x_2$ one has $f(x_1) \leq f(x_2)$. If the order "≤" in the definition of monotonicity is replaced by the strict order "<", then one obtains a strictly monotonically increasing function. Note that $f'(x_t) = 0$ only at those isolated points, rather than a continuous range, otherwise it violates the definition of monotonicity.*

From Fig.9(a), we should have

$$
\begin{aligned}
\dot{y}_t &= \frac{d\, f(x_t)}{dt} \\
&= \frac{d\, f(x_t)}{dx_t} \cdot \frac{d\, x_t}{dt} \\
&= f'(x_t) \cdot \frac{d\,(1 - \cos t)}{dt} \\
&= f'(x_t) \cdot \sin t \\
&= f'(x_t) \cdot \dot{x}_t. \qquad (6)
\end{aligned}
$$

$\dot{y}_t = f'(x_t) \cdot \dot{x}_t$ is clearly depicted in Fig.9(b). Actually, Eq.6 verifies the so-called conformality between the *x*-*y* plane and the $\dot{x} - \dot{y}$ plane: the line tangent $f'(x_t)$ to the $y_t = f(x_t)$ curve at any operating point $(x_t, y_t)$ in the *x*–*y* plane is equal to the (chord) slope of a straight line connecting the projected point $(\dot{x}_t, \dot{y}_t)$ to the origin in the $\dot{x} - \dot{y}$ plane [5].

Since $\dot{x}_t = \sin t = \sin(\pi - t)$, there must be a pair of corresponding time variables ($t$ and $\pi$–$t$) for each $\dot{x}_t$. Accordingly, $f'(x_t) = f'(1 - \cos t) \neq f'[1 - \cos(\pi - t)]$, which implies that $\dot{y}_t = f'(x_t) \cdot \dot{x}_t$ must take two different values for each $\dot{x}_t$. In other words, the curve in the $\dot{x} - \dot{y}$ plane in Fig.9(b) must be double-valued.

Observing Eq.6, we should obviously have $\dot{y}_t = f'(x_t) \cdot \dot{x}_t = 0$ when $\dot{x}_t = 0$. That is, the curve in the $\dot{x} - \dot{y}$ plane in Fig.9(b) must cross the origin.

Since $\dot{x}_t = \sin t$ and $-\dot{x}_t = -\sin t = \sin(-t)$, we should obtain

$$
\begin{aligned}
\dot{y}_t(t) &= f'(x_t) \cdot \dot{x}_t \\
&= f'(1 - \cos t) \cdot \sin t \\
&= -f'[1 - \cos(-t)] \cdot \sin(-t) \\
&= -f'[1 - \cos(-t)] \cdot (-\dot{x}_t) \\
&= -\dot{y}_t(-t), \qquad (7)
\end{aligned}
$$

which implies that the curve in the $\dot{x} - \dot{y}$ plane in Fig.9(b) must be anti-symmetric against the origin.

Furthermore, we can use differentiation-by-parts to obtain

$$
\begin{aligned}
\ddot{y}_t &= \frac{d\, \dot{y}_t}{dt} \\
&= \frac{d\, [f'(x_t) \cdot \sin t]}{dt} \\
&= \frac{d\, f'(x_t)}{dt} \cdot \sin t + f'(x_t) \cdot cost\ t \\
&= \frac{d\, f'(x_t)}{d\, x_t} \cdot \frac{d\, x_t}{dt} \cdot \sin t + f'(x_t) \cdot cost\ t \\
&= f''(x_t) \cdot \frac{d\,(1 - \cos t)}{dt} \cdot \sin t + f'(x_t) \cdot cost\ t \\
&= f''(x_t) \cdot sin^2 t + f'(x_t) \cdot cost\ t \\
&= f''(x_t) \cdot sin^2 t + f'(x_t) \cdot \ddot{x}_t \\
&= f''(1 - \cos t) \cdot sin^2 t + f'(1 - \cos t) \cdot \ddot{x}_t. \qquad (8)
\end{aligned}
$$

For each $\ddot{x}_t(t) = \cos t = \cos(-t)$, there must be a pair of corresponding time variables ($t$ and –$t$). Accordingly, we have

$$\ddot{y}_t = f''(1 - \cos t) \cdot sin^2 t + f'(1 - \cos t) \cdot \cos t$$

$$= f''[(1-\cos(-t)]\cdot sin^2(-t) + f'[(1-\cos(-t)]\cdot \cos(-t), \quad (9)$$

which implies that the curve in the $\ddot{x}-\ddot{y}$ plane in Fig.9(c) must be single-valued.

When $\ddot{x}_t(t)=\cos t=0$, we have $t=\frac{\pi}{2},\frac{3\pi}{2},\ldots$. Accordingly, we should obviously obtain

$$\begin{aligned}\ddot{y}_t &= f''(1-\cos t)\cdot sin^2 t + f'(1-\cos t)\cdot \ddot{x}_t\\ &= f''(1)\cdot sin^2 t\\ &\neq 0 \end{aligned} \quad (10)$$

as $f''(1)\neq 0$ and $sin^2(\frac{\pi}{2},\frac{3\pi}{2},\ldots)\neq 0$. One may argue that, under a very special circumstance, the second-order derivate at the mid-point of the constitutive $f(x_t)$ curve could be zero. For example, if $f(x_t)=\frac{1}{2}x_t^2-\frac{1}{6}x_t^3$, we should have $f'(x_t)=x_t-\frac{1}{2}x_t^2$ and $f''(x_t)=1-x_t$, so that $f''(1)=1-1=0$, as shown in Fig.10. Nevertheless, in most cases, we still have $f''(1)\neq 0$. By proof-by-contradiction [18], even one single sample element with $f''(1)\neq 0$ is mathematically sufficient to prove that it is not passive.

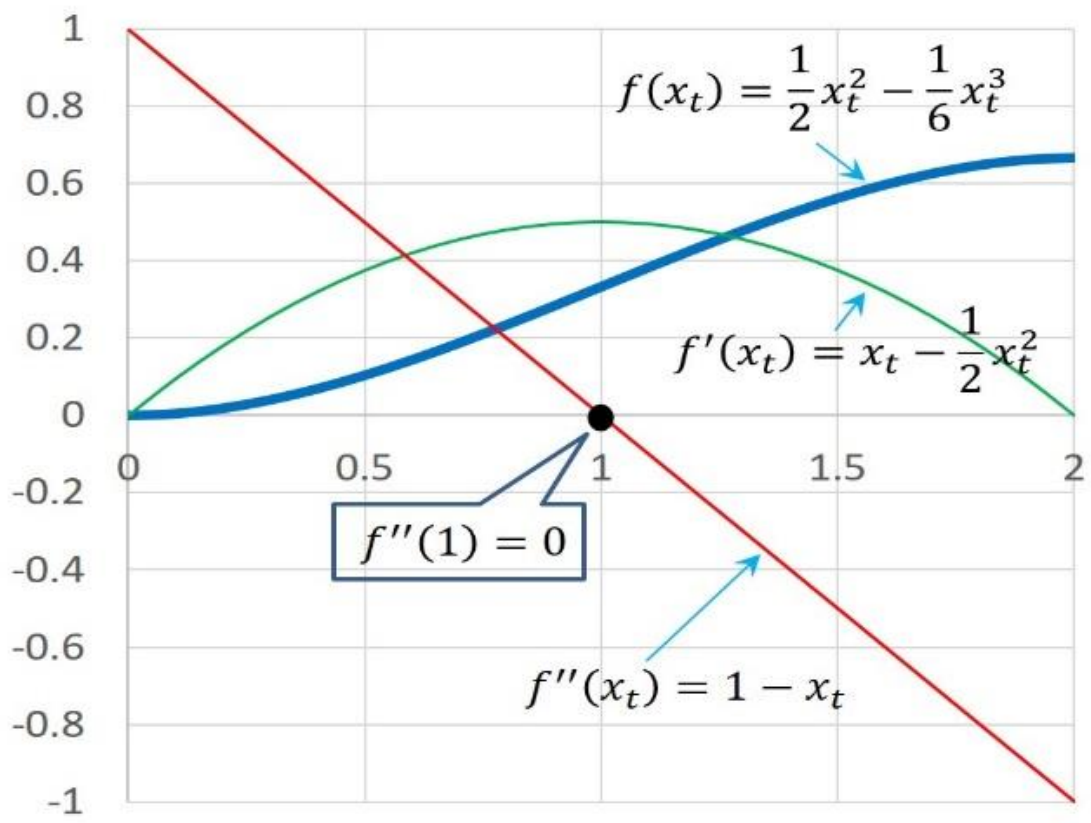

*Fig.10 Under a very special circumstance, the second-order derivate at the mid-point of the constitutive $f(x_t)$ curve could be zero.* In most cases, we still have $f''(1)\neq 0$.

In summary, the curve in the $\ddot{x}-\ddot{y}$ plane in Fig.9(c) must not go through the origin. That is, the corresponding circuit element is active. This analytical proof is completed.

## V. A Higher-order Ideal Mem-Inductor Degenerates into a Negative Nonlinear Inductor

In the same way to define a second-order ideal memristor (Fig4 or Fig.7), a second-order ideal mem-inductor should be characterized by a time-invariant ∫ρ-σ curve complying with the same set of four criteria as mentioned above.

After the first conformal differential transformation from the original ∫ρ-σ plane, the $\rho=\hat{\rho}(q)$ curve is a pinched Lissajous figure. Similarly, we can draw a horizontal secant to intersect the $\hat{\rho}(q)$ loop at the two intersecting points *A'* and *B'*. According to the mean value theorem, a point *C'* in (*A', B'*) must exist such that the tangent at *C'* is zero. After the second conformal differential transformation, *C'* is projected onto the *φ-i* plane as *C''*. This projected point *C''* is located on the *i* axis and apart from the origin ($C''\neq O''$), which indicates that there exists an internal current source; hence, a second-order mem-inductor is locally active.

The following activity theorem for a second-order or higher-order mem-inductor is therefore proven:

Theorem IV: Local Activity Theorem
A second-order or higher-order ideal mem-inductor should be locally active.

## VI. A Higher-order Ideal Mem-Capacitor Degenerates into a Negative Nonlinear Capacitor

In the same way to define a second-order ideal memristor (Fig4 or Fig.7), a second-order ideal mem-capacitor should be characterized by a time-invariant ρ-∫σ curve complying with the same set of four criteria as mentioned above.

After the first conformal differential transformation from the original ρ-∫σ plane, the $\varphi=\hat{\varphi}(\sigma)$ curve is a pinched Lissajous figure. In contrast, we now draw a vertical secant to intersect the $\hat{\varphi}(\sigma)$ loop at the two intersecting points *A'* and *B'*. According to the mean value theorem, there must exist a point *Q'* in (*A', B'*) such that the tangent at *Q'* is $\frac{\pi}{2}$. After the second conformal differential transformation, *Q'* is projected onto the *v-q* plane as *Q''*. This projected point *Q''* is located on the *v* axis and apart from the origin ($Q''\neq O''$), which indicates that there exists an internal voltage source; hence, a second-order mem-capacitor is locally active.

The following activity theorem for a second-order or higher-order mem-capacitor is therefore proven:

Theorem V: Local Activity Theorem
A second-order or higher-order ideal mem-capacitor should be locally active.

## VII. Conclusions and Discussions

As shown in Fig.11, our study can be simplified to an open mathematical/physical question:

Is a single-valued, nonlinear, continuously differentiable, and monotonically increasing constitutive curve locally active?

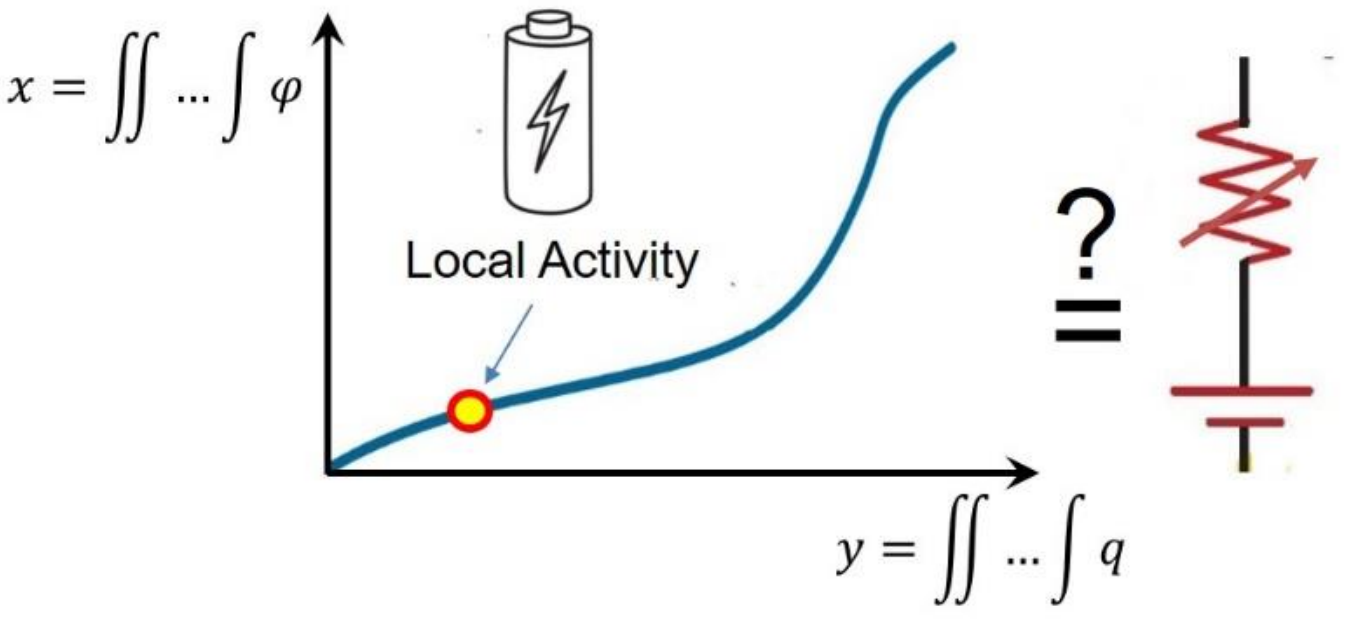

*Fig.11 An open mathematical/physical question: Is a single-valued, nonlinear, continuously differentiable, and monotonically increasing constitutive curve x-y locally active? Our theorems show that there must be at least one local active operating point on the curve.*

Based on both graphical and analytical proofs, our finding is that all higher-order passive memory circuit elements [memristor ($\alpha \leq -2$, $\beta \leq -2$), mem-inductor ($\alpha \leq -3$, $\beta \leq -2$), and mem-capacitor ($\alpha \leq -2$, $\beta \leq -3$)] must have an internal power source; hence, their passive versions do not exist in nature since they are bound to degenerate into zeroth-order negative nonlinear elements, as shown in Fig.11. Their active versions (with either an electric, optical, chemical, nuclear or biological power source) may still be found in nature (e.g., the first-order potassium memristor and the second-order sodium memristor in the Hodgkin-Huxley circuit [9][10]) or may be built as an artifact in the lab with the aid of active components (e.g., transistors or operational amplifiers) and power supplies. This phenomenon is similar to what was observed in Mendeleev's periodic table of chemical elements [19]: chemical elements with a higher atomic number are unstable and may decay radioactively into other chemical elements [12].

Consequently, the periodic table of the two-terminal passive circuit elements can be dramatically reduced to a six-pointed star comprising only six passive elements: resistor, inductor, capacitor, memristor, mem-inductor, and mem-capacitor, as shown in Fig.12. Such a bounded table is believed to be concise, mathematically sound and aesthetically beautiful, which may mark the finish of the hunt for missing higher-order (passive) circuit elements predicted by Chua in 1982 [1]. To date, all six elements in this six-pointed star have been fully studied: resistors [20], inductors [21], capacitors [21], memristors [3][4][5][22], mem-inductors [6][23], and mem-capacitors [6].

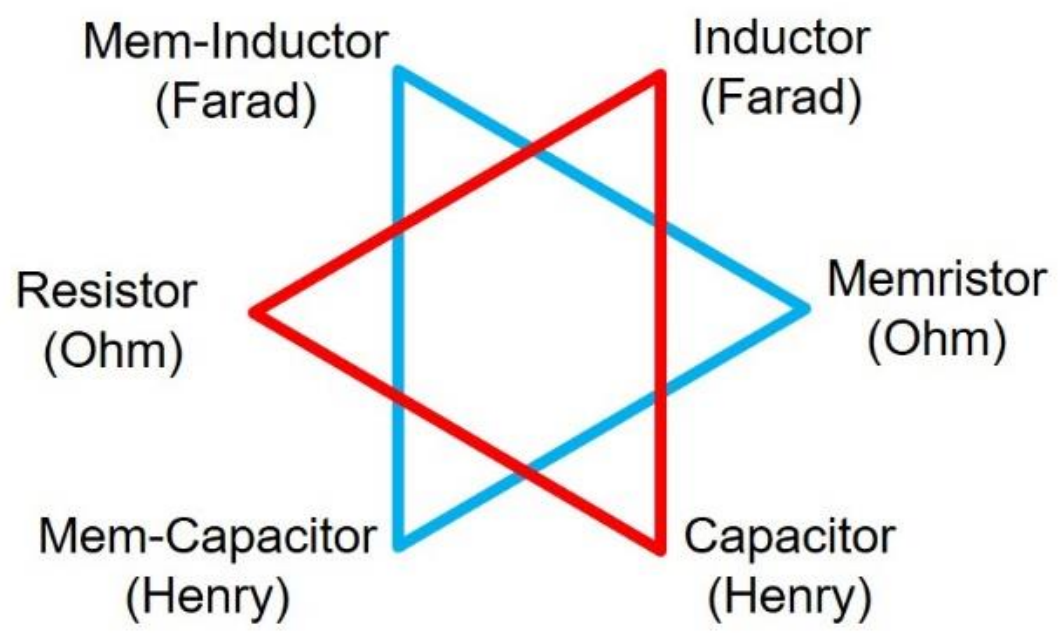

*Fig.12 A six-pointed star of all passive circuit elements. It is composed of a pair of face-to-face equilateral triangles, with their intersection as a regular hexagon. The first triangle includes the three traditional circuit elements (resistor, capacitor and inductor) [5][24] whereas the second triangle includes the three new but argumentative circuit elements since Chua called his predicted memristor a fourth circuit element [5][25]. The units in the diagram highlight the correspondence between the two triangles [25]. This new table is a big leap from an infinity in Chua's table (Fig.1) to a bound of 6 circuit elements only.*

Based on the fractional calculous [26][27][28][29][30], the table in Fig.12 can be extended to include more circuit elements: the intermediate case between the apexes (the integer-order circuit elements) represents fractional circuit elements, in which the interaction between the two constitutive attributes is fractional.

It is worth mentioning that different waveforms of the input signal may be used, which may result in different signature behaviors [3][4]. However, as mentioned in Section IV, it is proof-by-contradiction [18] that we used to prove our local activity theorems. That is, even one single counter-example (with any waveform of the input signal) is mathematically sufficient to prove that the corresponding element is not passive.

In "the case for rejecting the memristor as a fundamental circuit element" (the title of Abraham's 2018 paper [31]), a unique perspective is presented in the sense that that "charge" is the single fundamental electronic entity as well as its associated entity "voltage" (since separation of positive and negative charges builds up an electrical potential energy and a non-zero voltage appears). "Magnetic flux" and "current" are not thought by them to be fundamental any longer since both entities are merely a result of various states of motion of charge [31]. They used their charge/voltage-based periodic table to re-categorize the circuit elements and found that the memristor has no entry in their table [31]. We dispute it because motion of charge is not the only source to generate magnetic flux in all cases. As can be illustrated in our spin information storage work [32], magnetic flux may be caused by spontaneous magnetization or spin (that is independent from charge although we cannot separate the internal, intrinsic spin and charge of an electron). Even if it is "internal molecular currents" that generate the magnetization, it must be different or independent of an (externally applied) current [31]. In our opinion, flux (resulting from spontaneous magnetization or spin) is still fundamental as universally acknowledged by the whole magnetics community and memristor is fundamental in terms of linking the two fundamental physical attributes: charge and flux [25]. So is mem-capacitor or mem-inductor [25].

In all discussed elements, the pinched-loop (Fig.2-10 except Fig.5) is signature behavior observable in one plane. In the light of the findings reported in Ref. [33], obtaining this behavior is related to satisfying the phase and magnitude conditions of the theory of Lissajous figures which necessities a nonlinear device. Therefore, all these "higher-order" mem-devices boil down to the theory of Lissajous figures [33]. There is no memory in all of these "passive and nonlinear" devices because the memory is associated with the parasitic capacitors or inductors in the fabricated devices claiming to be "mem-devices" [34].

Interestingly, the leap from an infinity in Chua's table [1][3][24] to a bound of 6 circuit elements only in this six-pointed star may be analogous to a recent breakthrough in the bounded gap between primes in mathematics. In 2014, Zhang proved that the gap is bounded by an integer less than 70 million [35], which triggered a research heat in narrowing the prime gap. Over the past years, the bound has been impressively reduced to 246, 12 and 6 [36].